\documentclass[twocolumn]{aastex701}

\usepackage{CJK}
\usepackage{enumitem}
\usepackage{amsmath}
\usepackage{placeins}

\shortauthors{Zhang et al.}

\begin{document}
\begin{CJK*}{UTF8}{gbsn}

\title{Current Accretion State versus Black Hole Mass: Imprints on Multiphase Gas in IllustrisTNG}

\correspondingauthor{Xiaoxia Zhang}
\email{zhangxx@xmu.edu.cn}

\author[0000-0003-4832-9422]{Xiaoxia Zhang (张小霞)}  
\affiliation{Department of Astronomy, Xiamen University, Xiamen, Fujian 361005, People's Republic of China}
\email{}
\author[0000-0002-2853-3808]{Taotao Fang (方陶陶)}
\affiliation{Department of Astronomy, Xiamen University, Xiamen, Fujian 361005, People's Republic of China}
\email{}
\author[0000-0002-3940-2950]{Shulan Yan (鄢淑澜)}
\affiliation{Department of Astronomy, Xiamen University, Xiamen, Fujian 361005, People's Republic of China}
\email{}
\author[0000-0002-3462-4175]{Si-Yue Yu (余思悦)}
\affiliation{Department of Astronomy, Xiamen University, Xiamen, Fujian 361005, People's Republic of China}
\email{}
\author[0000-0003-3564-6437]{Feng Yuan (袁峰)}
\affiliation{Center for Astronomy and Astrophysics and Department of Physics, Fudan University, Shanghai 200438, People's Republic of China}
\email{}

\begin{abstract}
The coupling between AGN feedback and the multiphase circumgalactic medium (CGM) is central to understanding galaxy evolution. In a previous paper, we showed that the black hole mass (\(M_{\rm BH}\)) correlates negatively with cold gas mass in IllustrisTNG100. We now extend this to the Eddington ratio (\(\lambda_{\rm Edd}\)), which traces the current accretion state, using the same sample of 5089 central galaxies at \(z=0\) under the same group-specific partial correlation framework, controlling for stellar and dark matter halo mass. We find that \(\lambda_{\rm Edd}\) correlates positively with warm gas in star-forming galaxies (\(\rho \approx 0.6\)--\(0.7\)), persisting after jointly controlling for \(M_{\rm BH}\), the mass in the other non-star-forming gas phases, and specific star formation rate. The \(\lambda_{\rm Edd}\)--cold gas correlation instead reverses sign, from negative in star-forming galaxies (\(\rho \approx -0.32\)) to positive in quenched galaxies (\(\rho \approx 0.40\)). Controlling for \(M_{\rm BH}\) leaves this correlation essentially unchanged in all three galaxy types. Further controlling for the total gas mass in the other phases strengthens the star-forming correlation and weakens the quenched correlation to \(\rho \approx 0.21\), which remains significant. These results establish the Eddington ratio as a phase- and type-dependent tracer of the gas--black hole connection: the cold gas reservoir records the integrated black hole growth, the warm gas in star-forming galaxies carries the signature of the current accretion state, and the sign reversal of the cold-gas correlation marks the transition between the star-forming and quenched regimes.
\end{abstract}



\section{Introduction} 
\label{sec:intro}

Supermassive black holes inject vast amounts of energy into their surroundings through AGN feedback, shaping the evolution of their host galaxies \citep{2012ARA&A..50..455F, 2013ARA&A..51..511K}. This energy is deposited into the circumgalactic medium (CGM), the diffuse gas reservoir around the host dark matter halo. The CGM is inherently multiphase: it contains cold atomic and molecular gas traced by \ion{H}{1} 21\,cm and CO emission, respectively \citep[e.g.,][]{2017ApJS..233...22S, 2018MNRAS.476..875C}, cool and warm photoionized gas observed in UV absorption lines \citep[e.g.,][]{2011Sci...334..948T, 2014ApJ...792....8W}, and hot plasma that emits X-rays \citep[e.g.,][]{2019ApJ...884..169G}. Observational studies have typically focused on one phase at a time \citep[for a review, see][]{2017ARA&A..55..389T}, and how AGN feedback couples to the full range of gas phases remains poorly understood.

A central question is whether the black hole mass (\(M_{\rm BH}\)) or the instantaneous accretion rate (\(\dot{M}_{\rm BH}\)) is more closely tied to the gas content of galaxies. \(M_{\rm BH}\) serves as a proxy for the integrated history of black hole growth, whereas the Eddington ratio \(\lambda_{\rm Edd}\) reflects the current accretion state.  Studies of star formation quenching favor \(M_{\rm BH}\): it is a more robust predictor of whether a galaxy is quenched than \(\dot{M}_{\rm BH}\) or AGN luminosity \citep[e.g.,][]{2016ApJ...830L..12T, 2020MNRAS.493.1888T, 2022MNRAS.512.1052P, 2023ApJ...944..108B}. This has been interpreted as evidence that the integrated energy output of the AGN, rather than its current state, drives long-term gas depletion. Whether the relative associations of black hole mass and current accretion state depend on gas phase and galaxy population remains incompletely characterized. Observational studies typically examine \(M_{\rm BH}\) or \(\dot{M}_{\rm BH}\) in isolation, and often only one gas phase at a time. Cold atomic gas correlates more strongly with \(M_{\rm BH}\) than with stellar mass \citep[e.g.,][]{2024Natur.632.1009W}, while the cool ionized phase traced by \ion{C}{4} shows no significant correlation with \(M_{\rm BH}\) \citep[e.g.,][]{2024ApJ...970..115G}. AGN-driven neutral and molecular outflows, by contrast, have been linked to the instantaneous accretion state \citep[e.g.,][]{2014A&A...562A..21C}, suggesting that these two tracers may respond to different gas phases or to different stages of the feedback cycle.

Cosmological simulations allow both tracers to be measured for the same galaxies, which is difficult to achieve observationally. In IllustrisTNG (TNG), kinetic-mode feedback, which is tied to \(M_{\rm BH}\), has been shown to remove cold gas from massive halos and prevent further cooling \citep[e.g.,][]{2020MNRAS.491.4462D, 2020MNRAS.499..768Z}. It also leaves a clear imprint on the hot and cool phases of the CGM, producing X-ray emitting bubbles and regulating the thermodynamic state of the halo gas \citep[e.g.,][]{2021MNRAS.508.4667P, 2023MNRAS.518.5754R}. On larger scales, AGN feedback drives the X-ray luminosity dichotomy between star-forming and quenched galaxies in TNG \citep[e.g.,][]{2020MNRAS.494..549T}. 
 A recent random forest analysis of the same simulation found the black hole accretion rate, rather than \(M_{\rm BH}\), to be the most important feature for predicting \ion{H}{1} content \citep[e.g.,][]{2025ApJ...980..145L}. Cosmological simulations predict that high-Eddington-ratio AGN preferentially reside in gas-rich, star-forming galaxies on galactic scales \citep{2022MNRAS.514.2936W}, and that the CGM gas fraction at fixed halo mass correlates with \(M_{\rm BH}\) but not with the current accretion rate \citep{2019MNRAS.485.3783D}. Comparative studies across different simulations further show that the quantitative predictions for cold gas depend sensitively on the feedback implementation \citep[e.g.,][]{2020MNRAS.497..146D, 2021MNRAS.507.2383A}. These studies relate the current accretion state to gas content on galactic or halo scales, but none has resolved how the association varies across gas phases and galaxy populations. The present paper adds this phase- and population-resolved comparison.

In a previous paper \citep[][hereafter Paper I]{2026ApJ..1008..129Z}, we used the IllustrisTNG100 simulation to measure the partial correlation between \(M_{\rm BH}\) and multiphase gas, controlling for stellar and dark matter halo mass. We found that \(M_{\rm BH}\) correlates negatively with cold gas mass (\(\rho \approx -0.37\)), with the strongest signal on circumgalactic scales. Whether the current accretion state shows a distinct association with the CGM, and, if so, how that association differs from the \(M_{\rm BH}\) trend, was not addressed. Here we apply the same framework to the Eddington ratio \(\lambda_{\rm Edd}\), using the same galaxy sample and gas phase definitions, providing a direct comparison between \(M_{\rm BH}\) and \(\lambda_{\rm Edd}\) within the same galaxies and apertures.

The paper is organized as follows. Section~\ref{sec:method} describes the simulation, sample selection and definitions, and partial correlation methodology. Section~\ref{sec:res} presents our results, beginning with the full-sample comparison and the warm-gas residual analysis, followed by the dependence on galaxy type and robustness tests of the key correlations. Section~\ref{sec:discuss} discusses the physical implications, compares our findings with observations and previous work, and outlines caveats. We summarize our conclusions in Section~\ref{sec:sum}.

\section{Methodology}
\label{sec:method}

This work is based on the same IllustrisTNG100 simulation, galaxy sample, gas phase definitions, and partial correlation framework as in Paper I; we refer the reader to that paper for a complete description. We briefly summarize these elements in Sections~\ref{sec:sim}--\ref{sec:partial_corr} and highlight the key difference: the replacement of \(M_{\rm BH}\) with the Eddington ratio.

\subsection{The IllustrisTNG100 Simulation}
\label{sec:sim}

The IllustrisTNG project \citep{2018MNRAS.475..624N, 2018MNRAS.475..648P, 2018MNRAS.475..676S, 2018MNRAS.477.1206N, 2018MNRAS.480.5113M} is a suite of cosmological magneto-hydrodynamical simulations that model the formation and evolution of galaxies from $z=127$ to $z=0$. The TNG model includes prescriptions for gas cooling, star formation, stellar feedback, chemical enrichment, and supermassive black hole growth and feedback \citep{2018MNRAS.473.4077P}.

We use the flagship TNG100 run, which evolves a periodic volume of approximately $(110\,{\rm Mpc})^3$ with a dark matter mass resolution of $7.5 \times 10^6\,M_\odot$ and a baryonic mass resolution of $1.4 \times 10^6\,M_\odot$. We adopt the TNG cosmological parameters with $h = 0.6774$. 
Our analysis is based on the publicly available $z=0$ snapshot, the corresponding group and subhalo catalogs, and the SubLink merger trees \citep{2015MNRAS.449...49R}.

AGN feedback in TNG operates in two distinct modes \citep{2017MNRAS.465.3291W}. At high accretion rates, thermal energy is injected into the surrounding gas. At low accretion rates, the feedback switches to a kinetic mode in which momentum is injected via isotropic, stochastic winds. 
The transition between the two modes is set by a threshold $\chi$ in the Eddington ratio, which depends on the black hole mass:
\begin{eqnarray}
\chi = \min\left[0.002\left(\frac{M_{\rm BH}}{10^8\,M_\odot}\right)^{2}, 0.1\right].
\end{eqnarray}
This kinetic mode is the primary mechanism for quenching star formation in massive galaxies in TNG, as it drives high-velocity outflows that eject circumgalactic gas and prevent further cooling \citep{2020MNRAS.491.4462D, 2020MNRAS.499..768Z}.

\subsection{Sample Selection and Definitions}
\label{sec:sample}

\begin{figure}[tbp]
\centering
\includegraphics[width=0.47\textwidth]{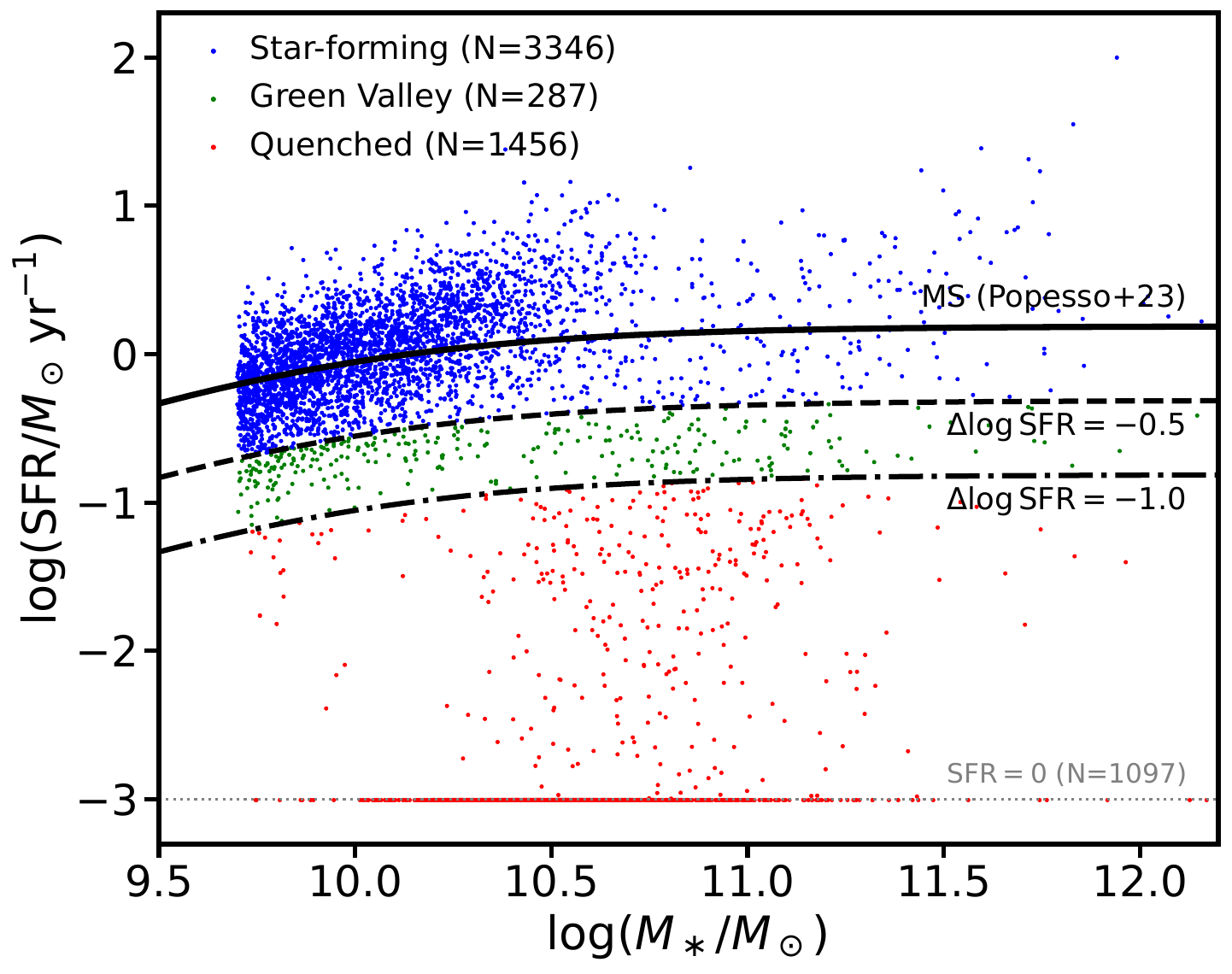}
\caption{Classification of the 5089 central galaxies in the SFR--$M_\ast$ plane. The solid curve is the $z=0$ star-forming main sequence of \citet{2023MNRAS.519.1526P}; the dashed and dot-dashed curves mark offsets of $-0.5$ and $-1.0$~dex. Colors show the three classes. Galaxies with $\mathrm{SFR}=0$ (1097 systems, below the resolution limit) are plotted at the bottom edge and classified as quenched.
\label{fig:class}}
\end{figure}

We select 5089 central galaxies at $z=0$ with stellar mass $M_* > 5 \times 10^9\,M_\odot$ and \(M_{\rm BH} > 0\). 
Following Paper~I, we classify galaxies by their offset from the $z=0$ star-forming main sequence of \citet{2023MNRAS.519.1526P}: star forming (within 0.5~dex of the main sequence), green valley (0.5--1.0~dex below), or quenched (more than 1.0~dex below). Galaxies with $\mathrm{SFR} = 0$ (1097 systems), whose star formation rate is below the resolution limit of the simulation, are assigned to the quenched class. The classification is shown in Figure~\ref{fig:class} and gives 3346 star-forming, 287 green-valley, and 1456 quenched galaxies.

We adopt the same definitions as in Paper~I. \(R_{200}\) encloses a mean density of 200 times the critical density. \(M_*\) is the stellar mass within twice the stellar half-mass radius, \(M_{\rm dm}\) is the total dark matter mass bound to the subhalo, and \(M_{\rm BH}\) is the sum of all black hole masses bound to the subhalo, dominated by the central black hole. The star formation rate is the sum of the star formation rates of all gas cells within twice the stellar half-mass radius.

Gas masses are computed from gas cells bound to the central subhalo, so gas bound to satellite subhalos is excluded even when it lies within \(R_{200}\) of the central galaxy. Gas temperature in TNG is computed from the internal energy and electron abundance of each cell. Star-forming cells are excluded, since TNG assigns them an effective temperature through the subgrid star-formation model rather than a thermodynamic value. The remaining non-star-forming gas is divided into four phases: cold ($T < 10^4\,{\rm K}$), cool ($10^4 \le T < 10^5\,{\rm K}$), warm ($10^5 \le T < 10^6\,{\rm K}$), and hot ($T \ge 10^6\,{\rm K}$). Gas masses are measured within three spherical apertures centered on the central galaxy, $0.03R_{200}$, $0.15R_{200}$, and $R_{200}$, chosen to probe the inner galaxy, the inner CGM, and the full virial region. Galaxies with zero gas mass in a given phase and aperture are omitted from the corresponding correlation analysis, since $\log M_{\rm gas}$ is undefined for them. The effective sample sizes are identical to those in Table~1 of Paper~I.

The key difference from Paper~I is the variable analyzed. Instead of \(M_{\rm BH}\), we use the Eddington ratio \(\lambda_{\rm Edd} = \dot{M}_{\rm BH} / \dot{M}_{\rm Edd}\), where \(\dot{M}_{\rm BH}\) is the instantaneous accretion rate of the central black hole and \(\dot{M}_{\rm Edd}\) is the Eddington accretion rate, both taken directly from the TNG snapshot at $z=0$. We verified that all selected galaxies have \(\dot{M}_{\rm BH} > 0\), so \(\lambda_{\rm Edd}\) is finite and positive for every galaxy analyzed here.

\subsection{Partial Correlation Analysis}
\label{sec:partial_corr}

Following Paper~I, we use a partial correlation analysis to isolate the intrinsic correlation between gas mass and \(\lambda_{\rm Edd}\), after removing the linear dependence on stellar and dark matter halo mass. For each galaxy type (star-forming, green valley, and quenched), we fit two linear regressions by ordinary least squares:
\begin{eqnarray}
\log M_{\rm gas} &=& a_1 \log M_* + b_1 \log M_{\rm dm} + c_1, \nonumber \\
\log \lambda_{\rm Edd} &=& a_2 \log M_* + b_2 \log M_{\rm dm} + c_2,
\end{eqnarray}
where \(M_{\rm gas}\) is the gas mass in a given phase and aperture, and \(M_*\) and \(M_{\rm dm}\) are the stellar and dark matter masses defined in Section~\ref{sec:sample}.
For each galaxy, the residual is the difference between the simulated value and the model prediction for its own galaxy type. The Spearman rank correlation between the gas-mass residuals and the \(\lambda_{\rm Edd}\) residuals gives the type-specific partial correlation coefficient \(\rho\). The full-sample partial correlation is obtained by pooling the type-specific residuals across all three types before computing the correlation. 

To test whether the key correlations are driven by other galaxy properties, we additionally control for \(M_{\rm BH}\), the total gas mass in the other phases, and specific star formation rate. When controlling for the total gas mass in the other phases, we exclude the target phase from the total to avoid a spurious part--whole correlation, so that the test probes whether the target phase carries information beyond the other phases. For the key results, we report the partial correlation coefficient, its analytic p-value, and its 95\% bootstrap confidence interval. We describe coefficients below \(|\rho| \approx 0.2\) as weak, even when formally significant.

\section{Results}
\label{sec:res}

\subsection{Full-Sample Comparison: \(\lambda_{\rm Edd}\) vs.\ \(M_{\rm BH}\)}
\label{sec:res_full}

\begin{figure*}[htbp]
\centering
\includegraphics[width=0.8\textwidth]{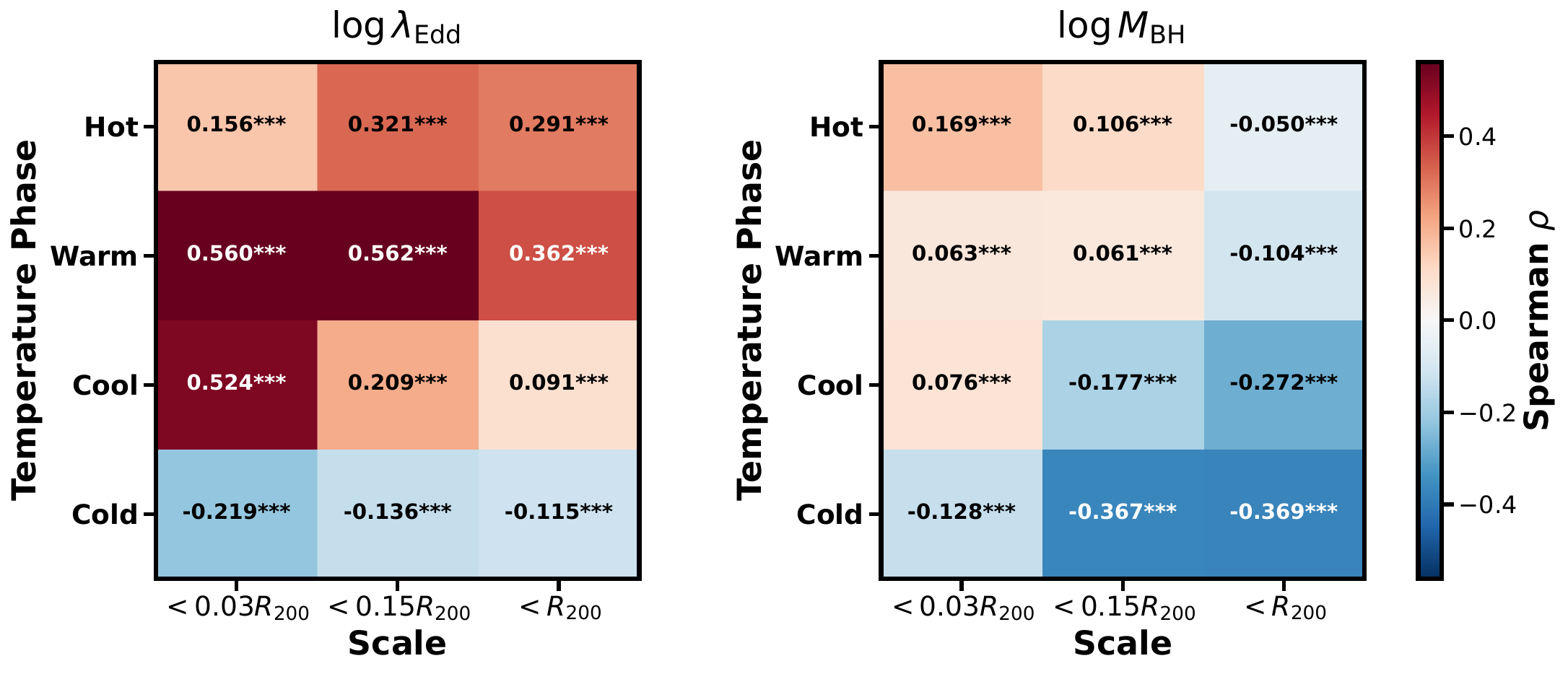}
\caption{Full-sample partial correlation between multiphase gas and the Eddington ratio (left) or black hole mass (right). Colors indicate Spearman's \(\rho\) after controlling for \(M_*\) and \(M_{\rm dm}\) via group-specific regressions. Numbers in cells are \(\rho\) values, with $^{*}$, $^{**}$, and $^{***}$ denoting $p<0.05$, $0.01$, and $0.001$, respectively.
\label{fig-1}}
\end{figure*}

Figure~\ref{fig-1} presents the full-sample partial correlation heatmaps for \(\lambda_{\rm Edd}\) (left panel) and \(M_{\rm BH}\) (right panel), after removing the linear effects of \(M_*\) and \(M_{\rm dm}\) via group-specific regressions. The \(M_{\rm BH}\) results, first presented in Paper~I, are shown here to facilitate a direct comparison with the Eddington ratio.

The two tracers show different patterns. The Eddington ratio correlates most strongly with warm gas at small and intermediate scales (\(\rho \approx 0.55\)), with a comparable signal for cool gas only at \(0.03R_{200}\). Hot gas shows a weak to moderate positive correlation at all scales, while cold gas shows a weak negative one. In the \(M_{\rm BH}\) panel,  the strongest signal is instead a moderate negative correlation with cold gas at large radii (\(\rho \approx -0.37\)), and the other phases show weak or moderate correlations.

The warm gas correlation is the strongest feature in the left panel, and we examine it in more detail in Section~\ref{sec:res_warm}.

\subsection{The Warm Gas Signal: Residual Analysis}
\label{sec:res_warm}

\begin{figure*}[htbp]
\centering
\includegraphics[width=0.8\textwidth]{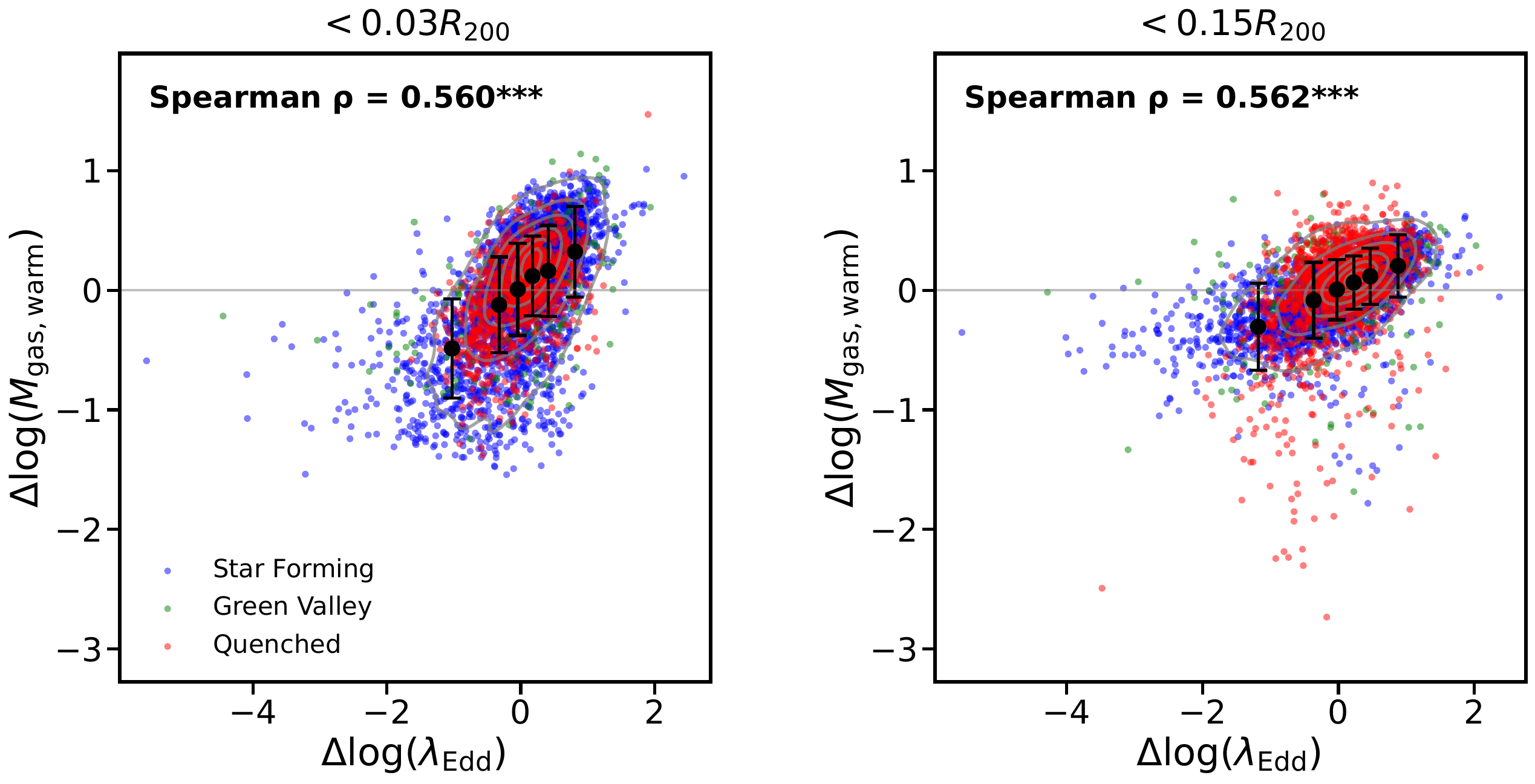}
\caption{Residual correlation between warm gas mass and \(\lambda_{\rm Edd}\) within \(0.03R_{200}\) (left) and \(0.15R_{200}\) (right), after removing the linear effects of \(M_*\) and \(M_{\rm dm}\). Colors indicate galaxy type: star-forming (blue), green valley (green), and quenched (red). Gray contours show the 2D density distribution. Black circles with error bars are binned means \(\pm1\sigma\) (six equal-frequency bins). The Spearman partial correlation coefficient is given in each panel.
\label{fig-2}}
\end{figure*}

Figure~\ref{fig-2} shows the residual correlation between warm gas and \(\lambda_{\rm Edd}\) within \(0.03R_{200}\) (left) and \(0.15R_{200}\) (right), where the partial correlation reaches \(\rho \approx 0.55\) at both scales. Each panel plots the gas mass residuals against the \(\lambda_{\rm Edd}\) residuals after removing the linear effects of \(M_*\) and \(M_{\rm dm}\), with galaxies colored by type. The binned means rise monotonically at both scales, supporting the use of the rank-based Spearman coefficient.

Star-forming galaxies span a wide range of \(\lambda_{\rm Edd}\) residuals and define the overall positive trend at both scales. Quenched galaxies occupy a narrower range of \(\lambda_{\rm Edd}\) residuals and are confined to the center of the distribution. Within \(0.03R_{200}\), quenched galaxies follow the same upward trend as star-forming galaxies with similar scatter, indicating that even at the low accretion rates typical of quenched systems,  the current accretion state is associated with the warm gas in the inner galaxy.

Within \(0.15R_{200}\), the slope flattens and the scatter among quenched galaxies increases. Both populations show a tail of galaxies with lower warm gas mass than expected, and the tail is more prominent among quenched galaxies.

\subsection{Galaxy Type Dependence}
\label{sec:res_type}

\begin{figure*}[tbp]
\centering
\includegraphics[width=0.98\textwidth]{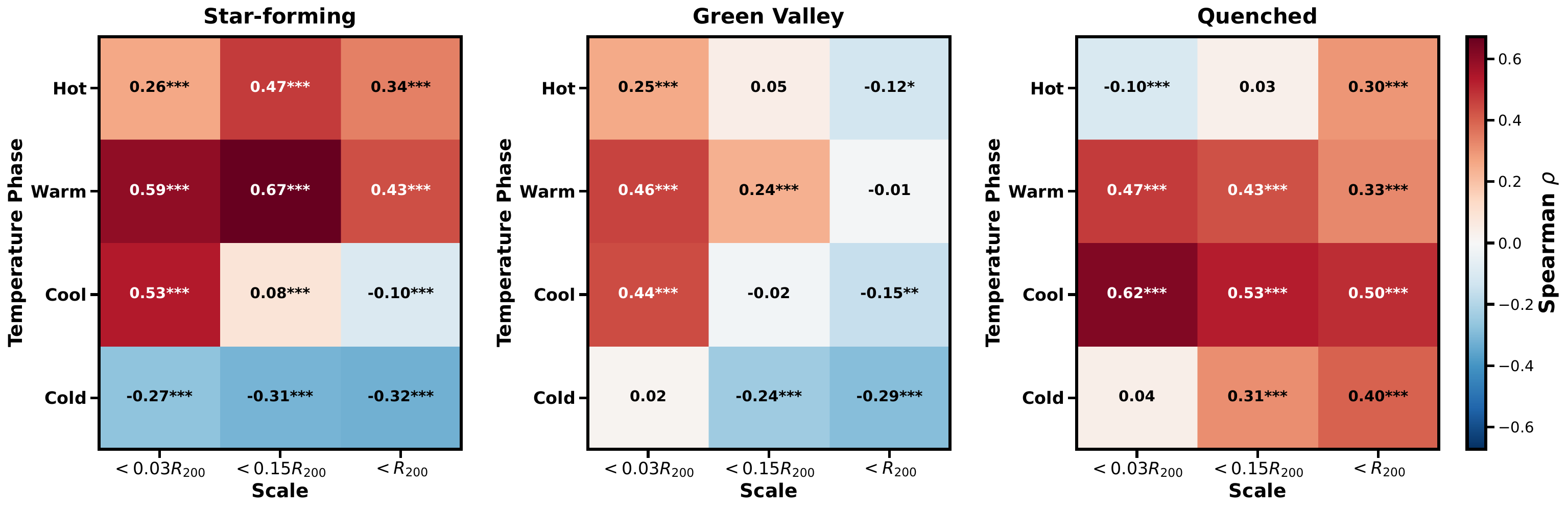}
\caption{Partial correlation between \(\lambda_{\rm Edd}\) and multiphase gas for star-forming (left), green valley (middle), and quenched (right) galaxies, after controlling for \(M_*\) and \(M_{\rm dm}\). Colors and significance levels follow the convention of Figure~\ref{fig-1}.
\label{fig-3}}
\end{figure*}

Figure~\ref{fig-3} splits the partial correlation between \(\lambda_{\rm Edd}\) and multiphase gas by galaxy type, revealing distinct patterns for star-forming, green valley, and quenched galaxies. The full-sample results in Figure~\ref{fig-1} therefore reflect a mixture of populations with different behavior.

In star-forming galaxies, \(\lambda_{\rm Edd}\) correlates most strongly with warm gas at small and intermediate scales (\(\rho \approx 0.6\)--\(0.7\)) and with cool gas at the innermost scale (\(\rho \approx 0.5\)). Both decline at the virial radius. Cold gas shows a moderate negative correlation at all three scales, with \(\rho \approx -0.3\).

Green valley galaxies show a mixed pattern. Warm and cool gas exhibit positive correlations at small radii, and cold gas is negatively correlated with \(\lambda_{\rm Edd}\) at \(0.15R_{200}\) and \(R_{200}\) (\(\rho \approx -0.25\) to \(-0.3\)). The correlations in this panel are generally less significant than in the star-forming panel, due to the small sample size (\(N=287\)).

Quenched galaxies present a different pattern. The cold gas correlation reverses sign, becoming positive at \(0.15R_{200}\) and \(R_{200}\) (\(\rho \approx 0.3\)--\(0.4\)). Warm and cool gas both remain positively correlated with \(\lambda_{\rm Edd}\) across all three scales, with cool gas showing the stronger correlation. Cool gas is thus the most robust \(\lambda_{\rm Edd}\) signal in quenched galaxies, while the warm gas correlation is weaker than in star-forming systems.

\begin{figure}[tbp]
\centering
\includegraphics[width=0.47\textwidth]{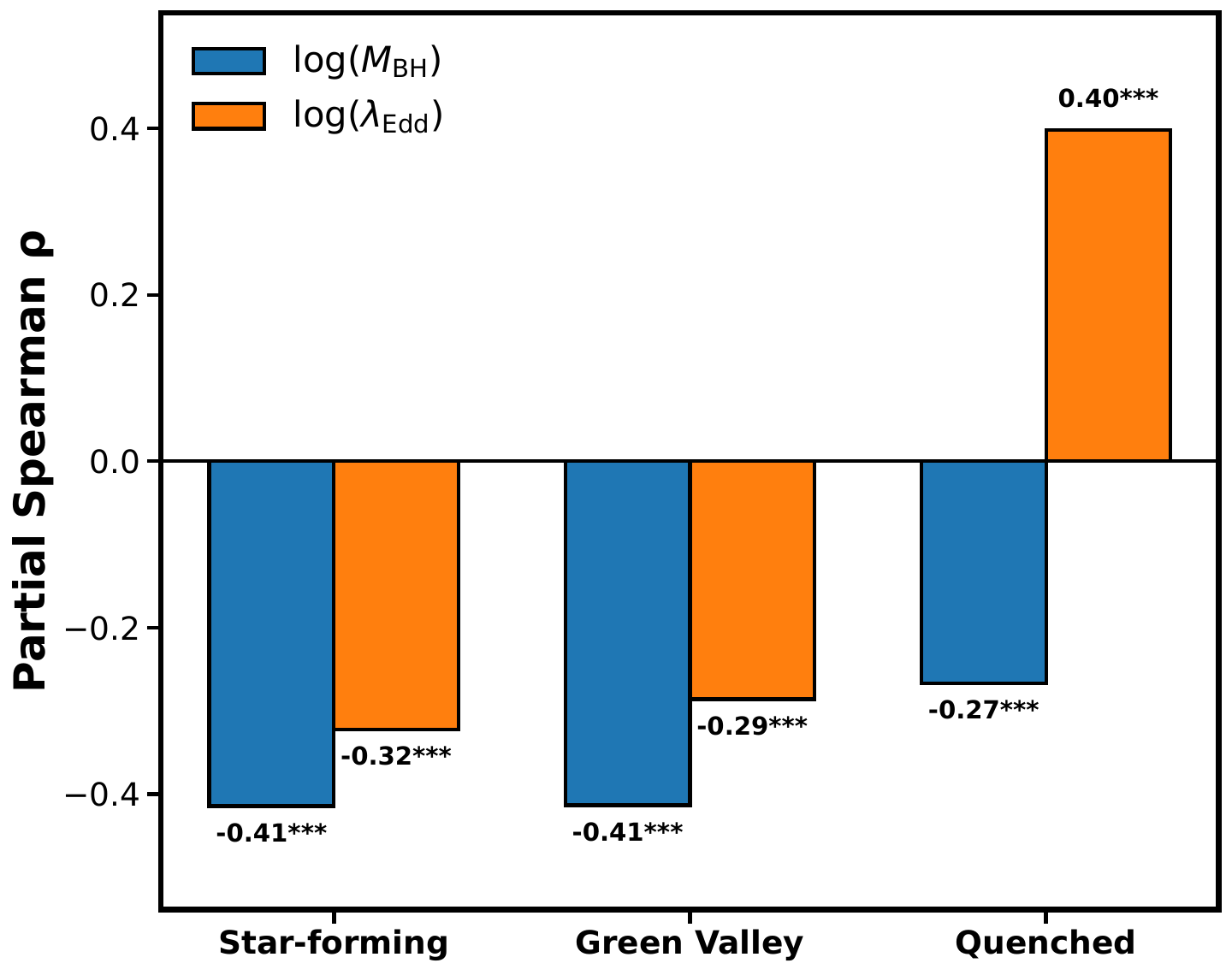}
\caption{Partial Spearman correlation coefficients between cold gas mass within \(R_{200}\) and \(M_{\rm BH}\) (blue) or \(\lambda_{\rm Edd}\) (orange) for star-forming, green valley, and quenched galaxies, after controlling for \(M_*\) and \(M_{\rm dm}\). Significance levels follow the convention of Figure~\ref{fig-1}.
\label{fig-4}}
\end{figure}

Figure~\ref{fig-4} compares the partial correlation coefficients for cold gas within \(R_{200}\) between the two tracers, making the sign reversal between star-forming and quenched galaxies explicit. For \(M_{\rm BH}\), the correlation with cold gas is negative across all three galaxy types (Paper~I). For \(\lambda_{\rm Edd}\), the correlation reverses sign, from negative in star-forming galaxies to positive in quenched galaxies, with green valley galaxies showing a value close to the star-forming population.

The \(\lambda_{\rm Edd}\)--gas correlations therefore depend strongly on galaxy type, and the cold gas sign reversal between star-forming and quenched galaxies is the most distinctive feature of the type-resolved analysis. 
All significant correlations in Figure~\ref{fig-3} keep their signs when $\lambda_{\rm Edd}$ is replaced by the black hole accretion rate $\dot{M}_{\rm BH}$ (Appendix~\ref{app:diag}, Figure~\ref{fig:mdot_heat}) or by averages of $\lambda_{\rm Edd}$ over the most recent 0.3 and 0.5\,Gyr (Figure~\ref{fig:windowavg}).
In Section~\ref{sec:res_robust} we further show that the sign reversal is robust to additional control variables, the choice of temperature threshold, and outlier removal.

\subsection{Robustness of Key Signals}
\label{sec:res_robust}

Two results warrant further investigation. The first is the sign reversal of the \(\lambda_{\rm Edd}\)--cold gas correlation between star-forming and quenched galaxies. Because \(\lambda_{\rm Edd} = \dot{M}_{\rm BH} / \dot{M}_{\rm Edd}\) and \(\dot{M}_{\rm Edd}\) scales with \(M_{\rm BH}\), the reversal may partly inherit its strength from the \(M_{\rm BH}\) dependence. The second is the positive correlation between \(\lambda_{\rm Edd}\) and warm/cool gas in star-forming galaxies, which may simply reflect that gas-rich systems contain more gas in every phase and sustain higher accretion rates. We test both possibilities by adding \(M_{\rm BH}\), the total gas mass in the other phases, and specific star formation rate as additional control variables, applied separately and jointly.

\begin{figure*}[tbp]
\centering
\includegraphics[width=0.98\textwidth]{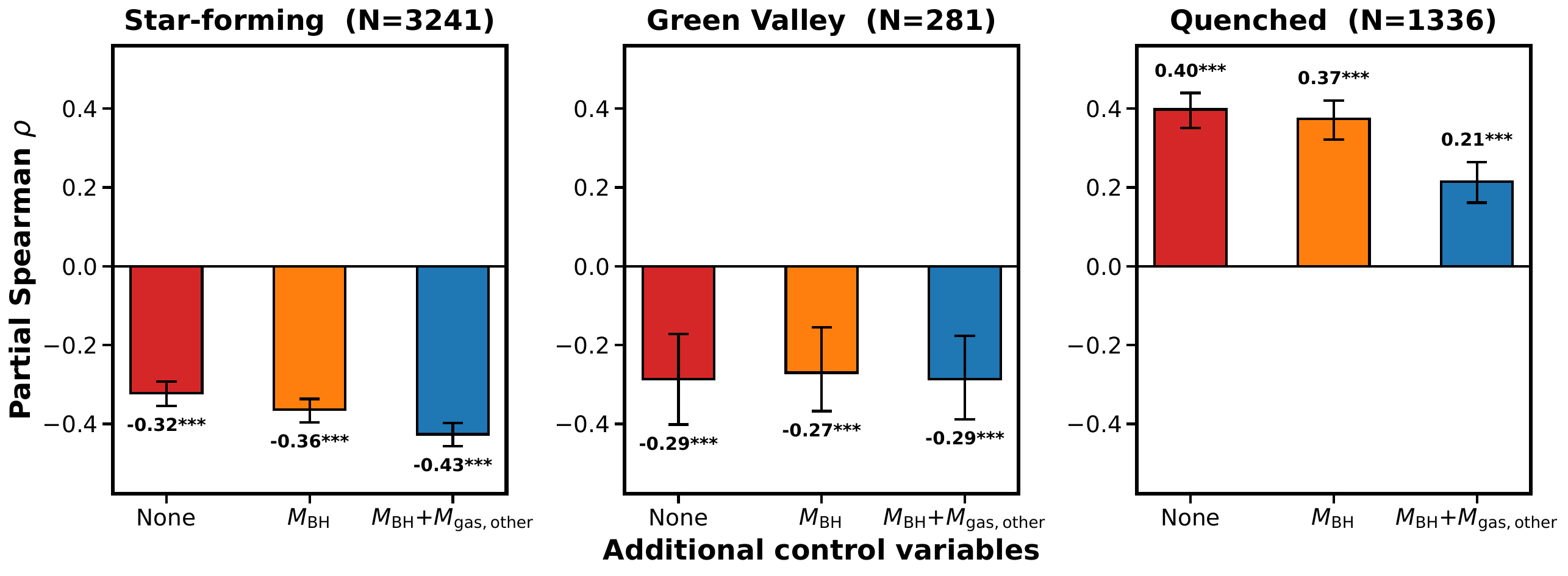}
\caption{Partial Spearman correlation between cold gas mass within \(R_{200}\) and \(\lambda_{\rm Edd}\) for star-forming, green valley, and quenched galaxies under three levels of additional control: none (red), \(M_{\rm BH}\) (orange), and \(M_{\rm BH}\) plus \(M_{\rm gas,other}\) (blue). Here \(M_{\rm gas,other}\) is the total gas mass in the non-star-forming phases other than cold gas, i.e., the cold phase is excluded from the control to avoid a part--whole correlation. Error bars show 95\% bootstrap confidence intervals.
\label{fig-5}}
\end{figure*}

Figure~\ref{fig-5} shows the cold gas correlation under three levels of control for all three galaxy types. Adding \(M_{\rm BH}\) as a control leaves the correlation essentially unchanged in all three galaxy types, and in star-forming galaxies it becomes slightly stronger (\(-0.32\) to \(-0.36\)). The sign reversal is therefore not inherited from the \(M_{\rm BH}\) dependence in the definition of \(\lambda_{\rm Edd}\). Further controlling for the total gas mass in the other phases strengthens the star-forming correlation to \(-0.43\) and reduces the quenched correlation to \(+0.21\), which remains significant.

\begin{figure*}[tbp]
\centering
\includegraphics[width=0.9\textwidth]{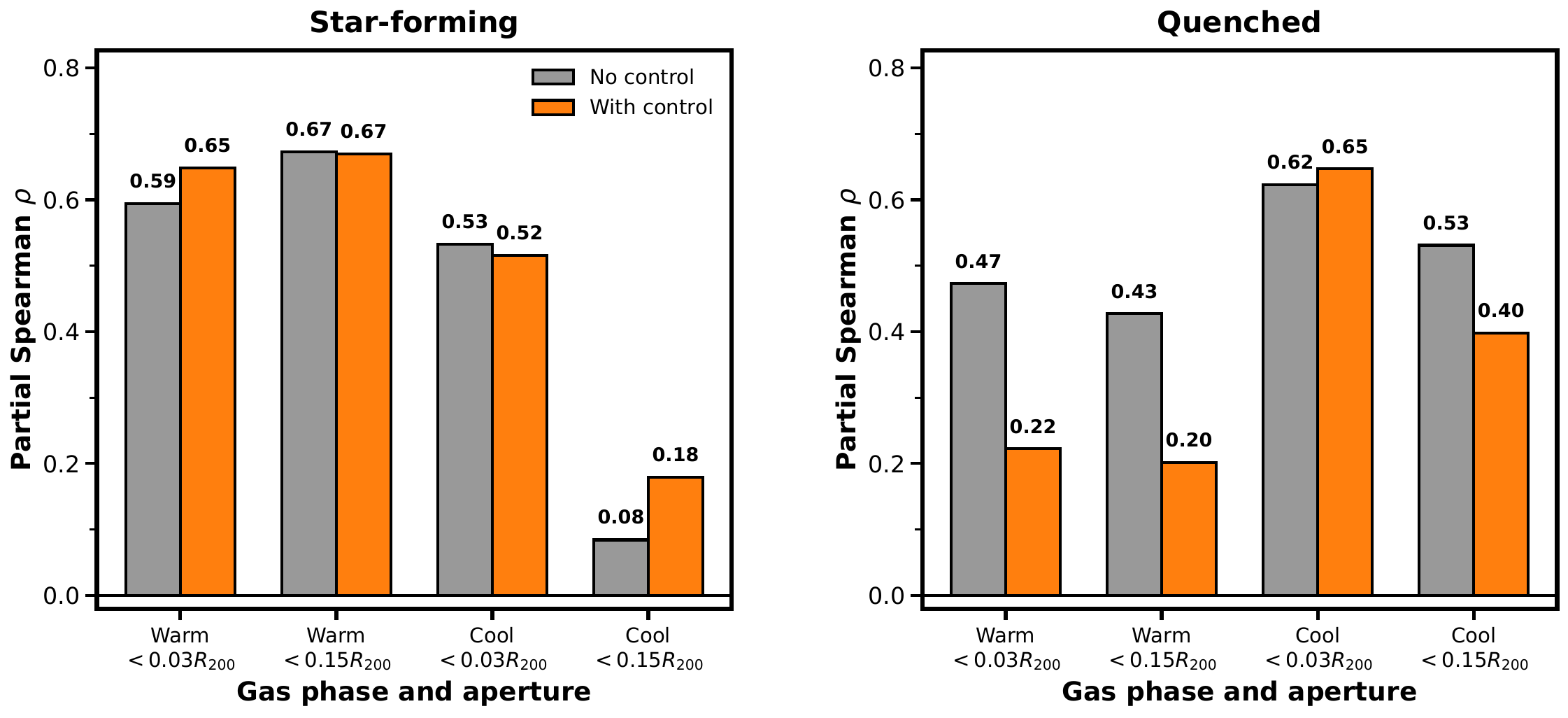}
\caption{Effect of controlling for the total gas mass in the other phases on the partial correlation between \(\lambda_{\rm Edd}\) and warm or cool gas. Each pair of bars compares no extra control (gray) with the additional control (orange). Left panel: star-forming galaxies. Right panel: quenched galaxies.
\label{fig-6}}
\end{figure*}

Figure~\ref{fig-6} focuses on the warm and cool phases at \(0.03R_{200}\) and \(0.15R_{200}\), which show the strongest correlations with \(\lambda_{\rm Edd}\). In star-forming galaxies, both correlations are almost unchanged or slightly enhanced by the additional control, and the cool gas correlation at \(0.15R_{200}\) stays weak with and without the control. In quenched galaxies, the warm gas correlation decreases by more than a factor of two once total gas is controlled, indicating that it largely reflects the common dependence on total gas supply. The cool gas correlation at \(0.03R_{200}\), by contrast, remains strong and increases slightly from 0.62 to 0.65.

For the warm gas signal in star-forming galaxies, we also perform a joint control test in which \(M_{\rm BH}\), the total gas mass in the other phases, and specific star formation rate are included simultaneously. The correlation remains essentially unchanged: \(\rho = 0.60\) (95\% bootstrap confidence interval: 0.57--0.62) at \(0.03R_{200}\) and \(\rho = 0.61\) (95\% confidence interval: 0.58--0.65) at \(0.15R_{200}\), both with \(p<0.001\). The warm gas association is therefore not driven by gas richness, black hole mass, or star formation activity, either individually or jointly.

The quenched cold gas reversal is also robust to the choice of temperature threshold. Redefining the cold phase as \(T < 10^{4.5}\)\,K or \(T < 10^{5}\)\,K preserves the sign and significance of the correlation in all three galaxy types. For quenched galaxies the reversal remains at \(\rho \approx 0.4\) and is insensitive to the threshold, whereas for star-forming galaxies the negative correlation weakens to \(\rho \approx -0.17\) but remains highly significant. The star-forming anticorrelation is thus concentrated in the coldest gas, whereas the quenched correlation is insensitive to the threshold.

The quenched reversal is also robust to the removal of outliers. Excluding values outside the 5th--95th percentile range of both \(\lambda_{\rm Edd}\) and cold gas mass within \(R_{200}\) leaves \(\rho = 0.32\), still positive and significant. 
Finally, a non-parametric random-forest test confirms that the black hole variables carry predictive power for the gas content beyond the galaxy and halo masses, independently of the linear-control assumption of the main analysis (Appendix~\ref{app:rf}, Figure~\ref{fig:rf}).

\section{Discussion}
\label{sec:discuss}

\subsection{Physical Interpretation}
\label{sec:discuss_physical}

The contrasting associations of \(M_{\rm BH}\) and \(\lambda_{\rm Edd}\) with the multiphase CGM reflect two different aspects of the coupled black-hole growth, fueling, and feedback cycle. \(M_{\rm BH}\) traces the integrated energy output of the black hole over cosmic time, which in TNG is closely tied to the cold gas content of the circumgalactic reservoir \citep{2020MNRAS.491.4462D, 2020MNRAS.499..768Z}. \(\lambda_{\rm Edd}\) reflects the current accretion state and correlates most strongly with the warm phase in star-forming galaxies. The two quantities therefore provide complementary views of a coupled fueling--feedback cycle, rather than two independent measures of feedback power.

The warm gas correlation in star-forming galaxies survives separate controls for mass in the other gas phases and specific star formation rate, and it also survives when these are applied jointly with \(M_{\rm BH}\) (Section~\ref{sec:res_robust}). This points to an association between the current accretion state and the warm gas that is not captured by gas richness, star formation activity, or black hole mass alone. The warm phase lies near the temperature range where radiative cooling is most efficient, so this association may reflect gas cooling toward the black hole, gas heated by feedback, or a combination of the two.

In the full sample, \(\lambda_{\rm Edd}\) correlates only weakly with cold gas. This weak correlation arises because the negative correlation in star-forming galaxies and the positive correlation in quenched galaxies partially cancel when the two populations are pooled, as shown in Figures~\ref{fig-3} and \ref{fig-4}. In addition, the instantaneous \(\lambda_{\rm Edd}\) probes only a brief moment of the accretion history: the accretion rate varies on \(10^5\)--\(10^6\)\,yr timescales \citep{2026arXiv260825341Z}, whereas the cold gas reservoir records feedback integrated over much longer, Gyr timescales (Paper~I).

The reversal of the \(\lambda_{\rm Edd}\)--cold gas correlation between star-forming and quenched galaxies shows the opposite regime. There, past feedback has already removed most of the cold gas \citep{2020MNRAS.493.1888T}, and the positive correlation is consistent with a picture in which the residual cold gas is the fuel for the current accretion. This reversal survives control for \(M_{\rm BH}\) and for the total gas mass in the other phases (Section~\ref{sec:res_robust}), and in star-forming galaxies the anticorrelation becomes stronger after these controls. The sign change is therefore not an artifact of the \(M_{\rm BH}\) normalization in the definition of \(\lambda_{\rm Edd}\). \(M_{\rm BH}\), by contrast, correlates negatively with cold gas across all galaxy types (Paper~I).

\subsection{Comparison with Observations and Previous Work}
\label{sec:discuss_comp}

Our results identify a strong \(\lambda_{\rm Edd}\)--cool gas correlation at \(0.03R_{200}\) in star-forming galaxies, which essentially disappears in the larger, cumulative apertures. The correlation with \(M_{\rm BH}\) is instead weak at \(0.15R_{200}\) and only modest at \(R_{200}\). This weakness at CGM scales is consistent with \citet{2024ApJ...970..115G}, who found that \ion{C}{4} absorption over 25--130 kpc correlates with the specific star formation rate but not with \(M_{\rm BH}\); the C\,\textsc{iv} absorption in such CGM observations typically traces photoionized gas at \(T \lesssim 10^5\)\,K, corresponding to our cool phase. The inner-scale \(\lambda_{\rm Edd}\) signal has no direct observational counterpart yet; it predicts a positive association between the current accretion state and cool gas at galactocentric radii of a few to $\sim 10$ kpc in star-forming galaxies, which could be tested with absorption studies and spatially resolved or down-the-barrel emission diagnostics.

For the cold phase, the roles reverse: \(\lambda_{\rm Edd}\) shows only a weak net correlation with cold gas in the full sample, while \(M_{\rm BH}\) anticorrelates with cold gas across all galaxy types (Paper~I), consistent with cumulative feedback being the primary driver of cold gas depletion \citep[e.g.,][]{2022MNRAS.512.1052P, 2023ApJ...944..108B, 2024Natur.632.1009W}. For the hot phase, \(M_{\rm BH}\) shows only weak correlations in the full sample, consistent with X-ray observations indicating that the hot halo is governed primarily by the gravitational potential \citep[e.g.,][]{2018ApJ...852..131B, 2019ApJ...884..169G, 2019MNRAS.488L.134L, 2019ApJ...875..141P}, whereas \(\lambda_{\rm Edd}\) shows moderate positive correlations at the two larger apertures, peaking at \(0.15R_{200}\) in star-forming galaxies.

The relative importance of \(\lambda_{\rm Edd}\) and \(M_{\rm BH}\) depends on the phase being examined and on which of the two is controlled. This helps explain the apparent tension between \citet{2025ApJ...980..145L}, who found the accretion rate to be the most important predictor of \ion{H}{1} content, and earlier work emphasizing kinetic-mode feedback tied to \(M_{\rm BH}\) \citep[e.g.,][]{2020MNRAS.491.4462D, 2020MNRAS.499..768Z}. In EAGLE, the CGM gas fraction at fixed halo mass correlates with \(M_{\rm BH}\) but not with the instantaneous accretion rate \citep[e.g.,][]{2019MNRAS.485.3783D}, broadly consistent with our finding for cold gas but leaving open whether the \(\lambda_{\rm Edd}\)--warm association is present in a thermal-only feedback model.

\subsection{Caveats}
\label{sec:discuss_caveats}

Several limitations of the present analysis should be noted.

First, \(\lambda_{\rm Edd}\) is measured from a single snapshot, and the partial correlation framework cannot establish causal direction. The key signals are nonetheless unchanged when \(\lambda_{\rm Edd}\) is averaged over the past 0.3--0.5\,Gyr (Figure~\ref{fig:windowavg}), so they are not driven by the exact snapshot timing. In TNG, the accretion rate responds to surrounding gas conditions and also helps determine the feedback mode, so \(\lambda_{\rm Edd}\) reflects a coupled cycle of gas fueling and feedback. For the same reason, we do not attempt to separate the contributions of thermal and kinetic feedback, since the mode boundary itself depends on \(M_{\rm BH}\) and \(\lambda_{\rm Edd}\) \citep{2017MNRAS.465.3291W}.

Second, the cold gas phase is unevenly resolved across the radial range probed here. In the innermost aperture (\(0.03R_{200}\)), the cold gas accounts for only a small fraction of the total gas and may be dominated by gas near the TNG cooling floor at \(10^4\)\,K. Our cold gas results are based on the \(R_{200}\) aperture, where the cold phase is well populated. They are also robust to the temperature threshold (Section~\ref{sec:res_robust}).

Third, in star-forming galaxies, stellar feedback also heats the surrounding gas, and star formation and black hole accretion share the same cold gas reservoir, so these two channels cannot be fully separated. In quenched galaxies, where star formation is largely absent, the residual positive correlations cannot be explained by ongoing star formation alone, although the present analysis cannot distinguish black-hole fueling, feedback, and other correlated halo processes. Because the phase-dependent associations depend on how feedback is implemented, extending this analysis to other simulations will be essential for assessing their generality.


\section{Summary and Conclusions}
\label{sec:sum}

We have measured the partial correlations between multiphase gas mass and two black hole properties, the black hole mass (\(M_{\rm BH}\)) and the Eddington ratio (\(\lambda_{\rm Edd}\)), in IllustrisTNG100. Using the same sample of 5089 central galaxies at \(z=0\) and the same group-specific partial correlation framework as in Paper~I, we correlate each property with the mass of cold, cool, warm, and hot gas within \(0.03R_{200}\), \(0.15R_{200}\), and \(R_{200}\), after removing the linear dependence on stellar and dark matter halo mass. Our main conclusions are as follows.

\begin{enumerate}

\item \(\lambda_{\rm Edd}\) correlates positively with warm gas in star-forming galaxies (\(\rho \approx 0.6\)--\(0.7\)), with the strongest signal at small and intermediate scales. This association survives joint control for \(M_{\rm BH}\), the total gas mass in the other phases, and specific star formation rate. In quenched galaxies, the strongest \(\lambda_{\rm Edd}\)--gas association after these controls is with cool gas at small radii (\(\rho \approx 0.6\)), while the warm gas signal is largely absorbed by the gas mass in the other phases.

\item The \(\lambda_{\rm Edd}\)--cold gas correlation reverses sign, from negative in star-forming galaxies (\(\rho \approx -0.32\)) to positive in quenched galaxies (\(\rho \approx 0.40\)). The reversal survives control for \(M_{\rm BH}\) and for the total gas mass in the other phases: in star-forming galaxies the anticorrelation becomes slightly stronger, whereas the quenched correlation weakens to \(\rho \approx 0.21\) but remains significant. The sign change is therefore not inherited from the \(M_{\rm BH}\) normalization in the definition of \(\lambda_{\rm Edd}\).

\item In the full sample, \(\lambda_{\rm Edd}\) shows only a weak net correlation with cold gas, because the opposite-sign correlations in the two populations partially cancel. Cold gas is instead most strongly tied to \(M_{\rm BH}\), which anticorrelates with cold gas across all three galaxy types and reaches \(\rho \approx -0.37\) in the full sample (Paper~I), while its full-sample correlations with the other phases are weak to modest. The current accretion state and the accumulated black hole growth thus carry different information about the gas--black hole connection.

\end{enumerate}

The picture that emerges is that the gas--black hole connection varies with gas phase and galaxy type, and neither \(M_{\rm BH}\) nor \(\lambda_{\rm Edd}\) alone fully describes it. Extending the accretion-rate averaging beyond the 0.3--0.5\,Gyr windows tested here to the full accretion history along merger trees would connect the instantaneous and cumulative views and show whether the phase-dependent associations vary continuously with timescale. Applying the same framework to other cosmological simulations would test whether these associations are generic or specific to the TNG feedback model.

\begin{acknowledgments}
This work is supported by the National SKA Program of China No. 2025SKA0150103, National Natural Science Foundation of China under Grant Nos. 12550002, 12133008, 12221003, 11890692, 12233005. We acknowledge the science research grants from the China Manned Space Project with No. CMS-CSST-2021-A04 and No. CMS-CSST-2025-A10, the Fujian Provincial Natural Science Foundation of China (Grant No. 2024J08001), and the Natural Science Foundation of Xiamen, China (No. 3502Z202472007). This work was also supported by the Fundamental Research Funds for the Central Universities (No. 20720260040, 20720260230).
\end{acknowledgments}

\software{numpy \citep{2020Natur.585..357H},
          scipy \citep{2020NaMet..17..261V},
          matplotlib \citep{2007CSE.....9...90H},
          pandas \citep{2022zndo...3509134T},
          astropy \citep{2013A&A...558A..33A,2018AJ....156..123A,2022ApJ...935..167A}
          }

\appendix

\section{Robustness to the Accretion Diagnostic}
\label{app:diag}

\begin{figure*}[tbp]
\centering
\includegraphics[width=0.98\textwidth]{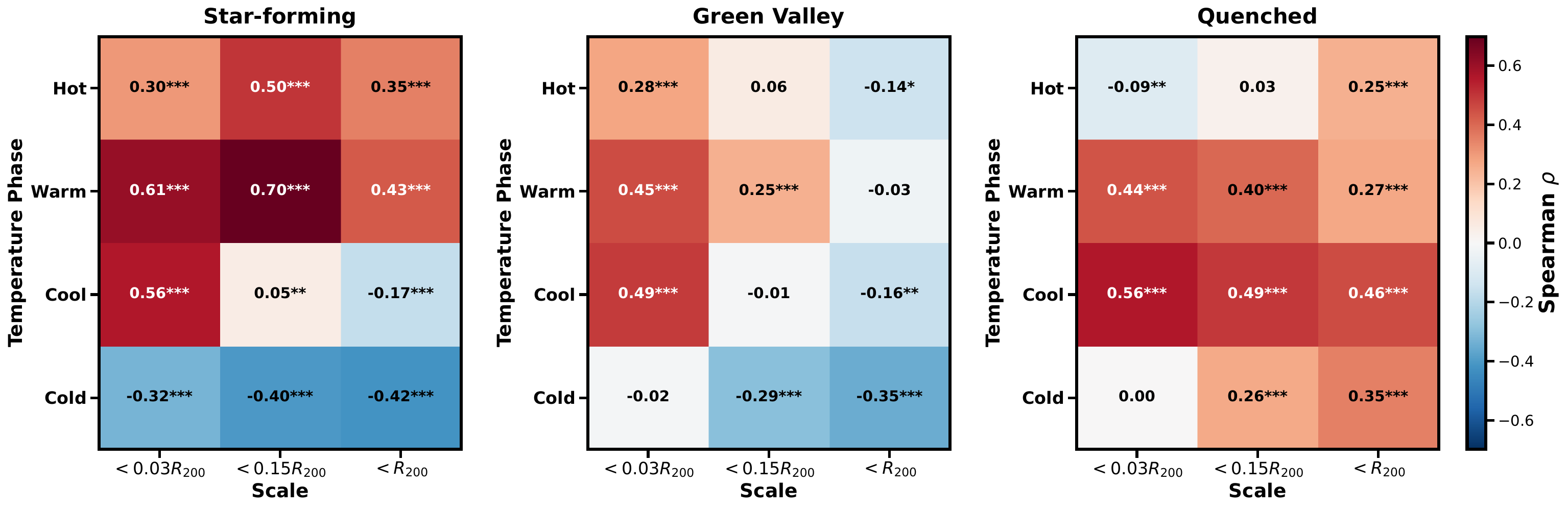}
\caption{Same as Figure~\ref{fig-3}, but with the black hole accretion rate $\dot{M}_{\rm BH}$ in place of $\lambda_{\rm Edd}$, for star-forming (left), green-valley (middle), and quenched (right) galaxies. Colors and significance levels follow the convention of Figure~\ref{fig-1}.
\label{fig:mdot_heat}}
\end{figure*}

We repeated the full correlation analysis with the black hole accretion rate $\dot{M}_{\rm BH}$ in place of the Eddington ratio. Unlike $\lambda_{\rm Edd}$, $\dot{M}_{\rm BH}$ is not normalized by the black hole mass, so the two diagnostics test different aspects of the accretion state. All significant correlations found with $\lambda_{\rm Edd}$ keep their signs with $\dot{M}_{\rm BH}$ (Figure~\ref{fig:mdot_heat}), including the cold-gas sign reversal between star-forming and quenched galaxies. The negative cold-gas correlations in star-forming and green-valley galaxies become somewhat stronger with $\dot{M}_{\rm BH}$, as expected once the black hole mass scaling is no longer divided out, while the positive correlation in quenched galaxies is slightly weaker. We keep $\lambda_{\rm Edd}$ as the primary diagnostic because it separates the accretion state from the mass scale.

\begin{figure*}[tbp]
\centering
\includegraphics[width=0.98\textwidth]{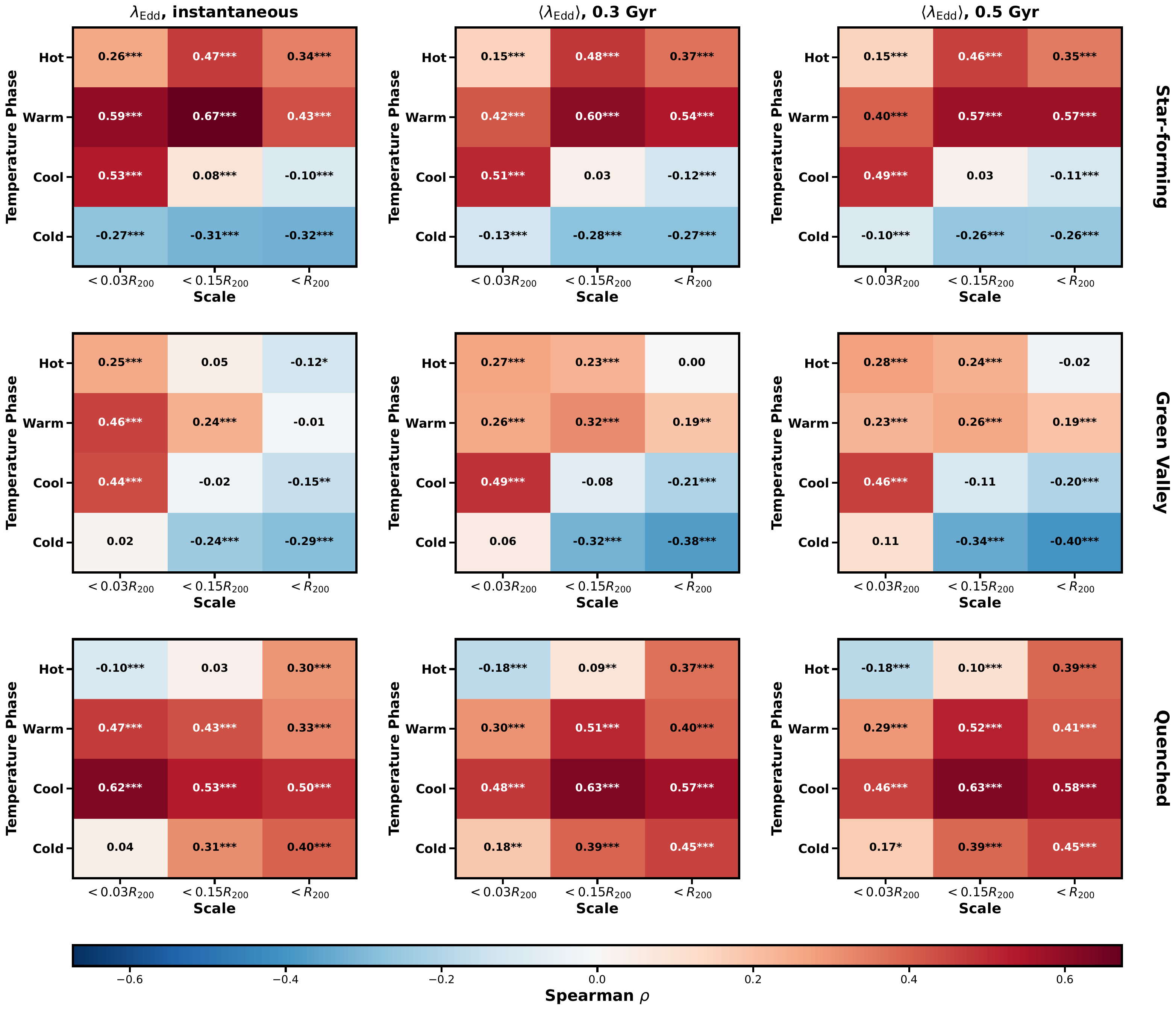}
\caption{Partial Spearman correlations between the gas phases and the Eddington ratio, for the instantaneous $\lambda_{\rm Edd}$ at $z=0$ (left column) and for averages over the most recent 0.3 and 0.5\,Gyr (middle and right columns), shown separately for star-forming (top), green-valley (middle), and quenched (bottom) galaxies. Correlations control for $\log M_\ast$ and $\log M_{\rm dm}$. Colors and significance levels follow the convention of Figure~\ref{fig-1}.
\label{fig:windowavg}}
\end{figure*}

As a second test, we replaced the instantaneous Eddington ratio with the accretion rate averaged over the snapshots of the SubLink main progenitor branch within the most recent 0.3 and 0.5\,Gyr, again normalized by the Eddington rate. The sign of every significant correlation is preserved under both averages, and the amplitudes change only modestly, typically by $\lesssim 0.1$ in $|\rho|$ (Figure~\ref{fig:windowavg}); in particular, the cold-gas sign reversal between star-forming and quenched galaxies is unchanged. A systematic study of longer averaging times will be presented in a companion paper (in preparation).

\section{Random-forest Validation}
\label{app:rf}

\begin{figure*}[tbp]
\centering
\includegraphics[width=0.6\textwidth]{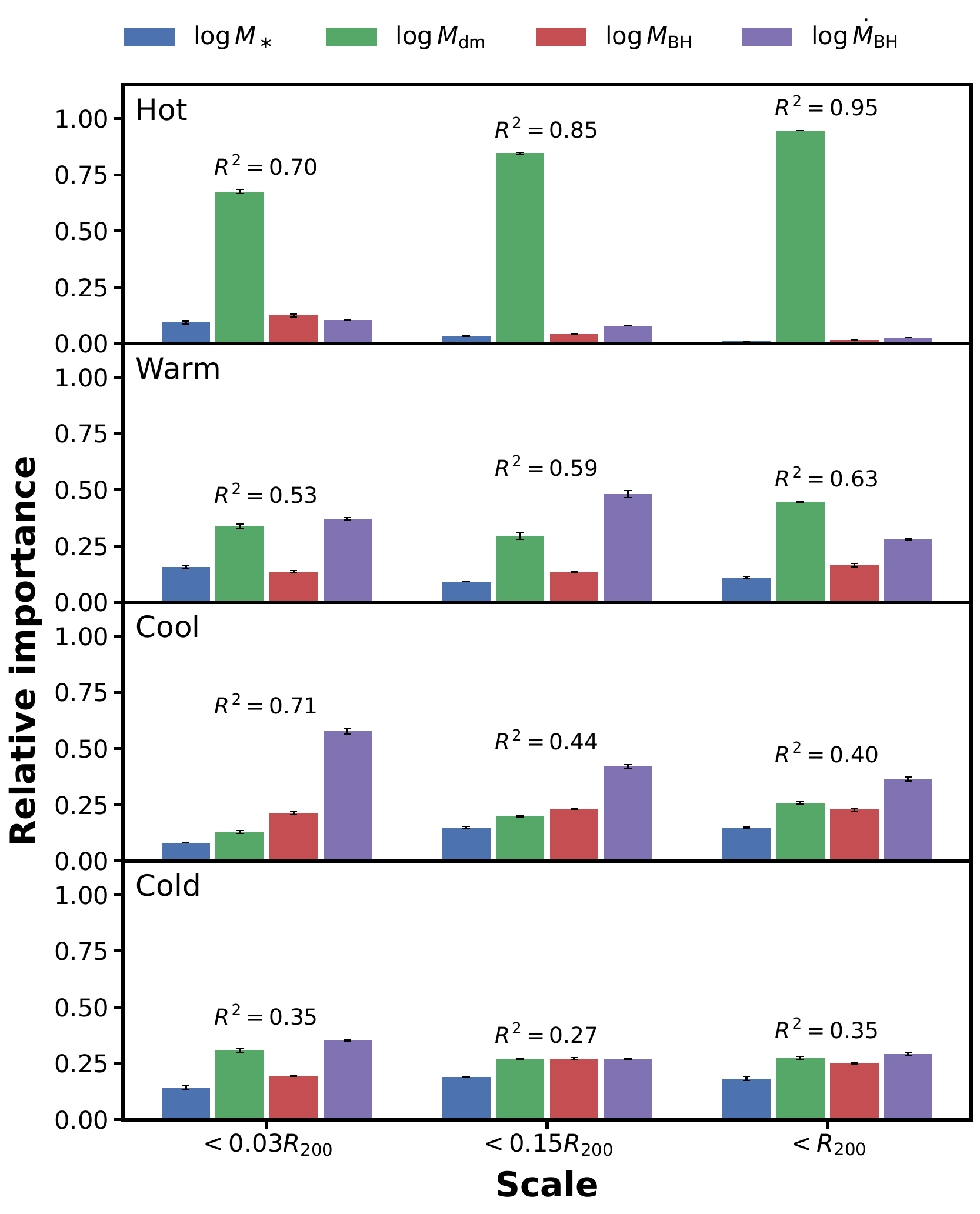}
\caption{Random-forest feature importance for predicting each gas phase within three apertures from $\log M_\ast$, $\log M_{\rm dm}$, $\log M_{\rm BH}$, and $\log\dot{M}_{\rm BH}$. Bars show the impurity-based share, averaged over three random seeds, with error bars giving the scatter between seeds. Permutation importance measured on the test sample confirms that the black hole variables retain substantial importance beyond the mass controls. The test-set $R^2$ (the fraction of the variance reproduced by the forest) is given above each aperture group.
\label{fig:rf}}
\end{figure*}

As a check that does not assume linear control, we trained random forest regressors \citep[\textsc{scikit-learn};][]{2011JMLR...12.2825P} to predict each gas-phase mass from $\log M_\ast$, $\log M_{\rm dm}$, $\log M_{\rm BH}$, and $\log\dot{M}_{\rm BH}$. We follow the ``All Parameter'' setup of \citet[][their Appendix~B]{2022A&A...659A.160B}, with all features available at each split. Each forest has 100 trees trained on 75\,\% of the sample and is scored on the remaining 25\,\% by $R^2$, the fraction of the variance in the test sample that the forest reproduces; we repeat the fit with three random seeds and show the scatter between seeds as error bars. We measure the importance of each feature in two ways, by its impurity-based share and by permutation on the test sample; a random control feature confirms that the permutation measure is the more reliable of the two. Both measures show that the black hole variables retain substantial importance beyond the mass controls (Figure~\ref{fig:rf}). The hot gas is dominated by the halo mass at all apertures. For the warm gas within $0.15\,R_{200}$, the accretion rate carries the largest importance share. For the cold gas within $R_{200}$, the two black hole variables together carry about half of the importance, comparable to the two mass controls. The black hole variables thus carry predictive power for the gas content beyond the galaxy and halo masses, independently of the linear-control assumption of the main analysis.

\bibliography{ms}{}
\bibliographystyle{aasjournal}

\end{CJK*}
\end{document}